\newif\ifpreprint
\preprintfalse 

\ifpreprint
\documentclass[journal=jpclcd,manuscript=letter]{achemso}
\else
\documentclass[journal=jpclcd,manuscript=letter,layout=twocolumn]{achemso}
\fi

\usepackage[T1]{fontenc} 

\usepackage{amsmath}
\usepackage{newtxtext,newtxmath}

\usepackage{pifont}
\usepackage{float}
\usepackage{graphicx}
\usepackage{dcolumn}
\usepackage{multirow}
\usepackage{xspace}
\usepackage{verbatim}
\usepackage[version=4]{mhchem} 
\usepackage{comment}
\usepackage{color,soul}

\usepackage{mathtools}
\usepackage[dvipsnames]{xcolor}
\usepackage{xspace}

\usepackage{graphicx,longtable,dcolumn,mhchem}
\usepackage{rotating,color}
\usepackage{lscape}
\usepackage{amsmath}
\usepackage{dsfont}
\usepackage{soul}
\usepackage{physics}

\newcommand{\ie}{\textit{i.e.}}
\newcommand{\eg}{\textit{e.g.}}

\definecolor{darkgreen}{HTML}{009900}
\usepackage[normalem]{ulem}
\newcommand{\titou}[1]{\textcolor{black}{#1}}

\newcommand{\ra}{\rightarrow}

\newcommand{\br}{\boldsymbol{r}}
\newcommand{\bx}{\boldsymbol{x}}

\newcommand{\DFT}{\text{DFT}}
\newcommand{\KS}{\text{KS}}
\newcommand{\BSE}{\text{BSE}}
\newcommand{\GW}{GW}
\newcommand{\XC}{\text{xc}}

\newcommand{\eps}{\varepsilon}

\newcommand{\HOMO}{\text{HOMO}}
\newcommand{\LUMO}{\text{LUMO}}
\newcommand{\Eg}{E_\text{g}}
\newcommand{\EgFun}{\Eg^\text{fund}}
\newcommand{\EgOpt}{\Eg^\text{opt}}
\newcommand{\EB}{E_B}

\newcommand{\Nel}{N}
\newcommand{\Norb}{N_\text{orb}}

\usepackage[colorlinks = true,
            linkcolor = blue,
            urlcolor  = black,
            citecolor = blue,
            anchorcolor = black]{hyperref}
\definecolor{goodorange}{RGB}{225,125,0}
\definecolor{goodgreen}{RGB}{5,130,5}
\definecolor{goodred}{RGB}{220,50,25}
\definecolor{goodblue}{RGB}{30,144,255}

\newcommand{\note}[2]{
\ifthenelse{\equal{#1}{F}}{
\colorbox{goodorange}{\textcolor{white}{\footnotesize \fontfamily{phv}\selectfont #1}}
    \textcolor{goodorange}{{\footnotesize \fontfamily{phv}\selectfont #2}}\xspace
}{}
\ifthenelse{\equal{#1}{R}}{
\colorbox{goodred}{\textcolor{white}{\footnotesize \fontfamily{phv}\selectfont #1}}
    \textcolor{goodred}{{\footnotesize \fontfamily{phv}\selectfont #2}}\xspace
}{}
\ifthenelse{\equal{#1}{N}}{
\colorbox{goodgreen}{\textcolor{white}{\footnotesize \fontfamily{phv}\selectfont #1}}
    \textcolor{goodgreen}{{\footnotesize \fontfamily{phv}\selectfont #2}}\xspace
}{}
\ifthenelse{\equal{#1}{M}}{
\colorbox{goodblue}{\textcolor{white}{\footnotesize \fontfamily{phv}\selectfont #1}}
    \textcolor{goodblue}{{\footnotesize \fontfamily{phv}\selectfont #2}}\xspace
}{}
}

\usepackage{titlesec}

\usepackage[fontsize=11pt]{scrextend}
\titleformat{\section}
{\normalfont\sffamily\bfseries\color{Blue}}
{\thesection.}{0.25em}{\uppercase}

\titleformat{\subsection}[runin]
{\normalfont\sffamily\bfseries}
{\thesubsection}{0.25em}{}[.\;\;]

\titleformat{\suppinfo}
{\normalfont\sffamily\bfseries}
{\thesubsection}{0.25em}{}

\titlespacing*{\section}{0pt}{0.5\baselineskip}{0.01\baselineskip}
\titlespacing*{\subsection}{0pt}{0.125\baselineskip}{0.01\baselineskip}

\author{Xavier Blase}
	\email{xavier.blase@neel.cnrs.fr}
	\affiliation[NEEL, Grenoble]{Universit\'e Grenoble Alpes, CNRS, Institut NEEL, F-38042 Grenoble, France}
\author{Ivan Duchemin}
	\affiliation[CEA, Grenoble]{Universit\'e Grenoble Alpes, CEA, IRIG-MEM-L Sim, 38054 Grenoble, France}
\author{Denis Jacquemin}
	\affiliation[CEISAM, Nantes]{Universit\'e de Nantes, CNRS,  CEISAM UMR 6230, F-44000 Nantes, France}
\author{Pierre-Fran\c{c}ois Loos}
	\email{loos@irsamc.ups-tlse.fr}
	\affiliation[LCPQ, Toulouse]{Laboratoire de Chimie et Physique Quantiques, Universit\'e de Toulouse, CNRS, UPS, France}

\let\oldmaketitle\maketitle
\let\maketitle\relax
	\title{The Bethe-Salpeter Equation Formalism: \\ From Physics to Chemistry} 
\date{\today}

\begin{tocentry}
	\vspace{1cm}
	\centering
	\includegraphics[width=\textwidth]{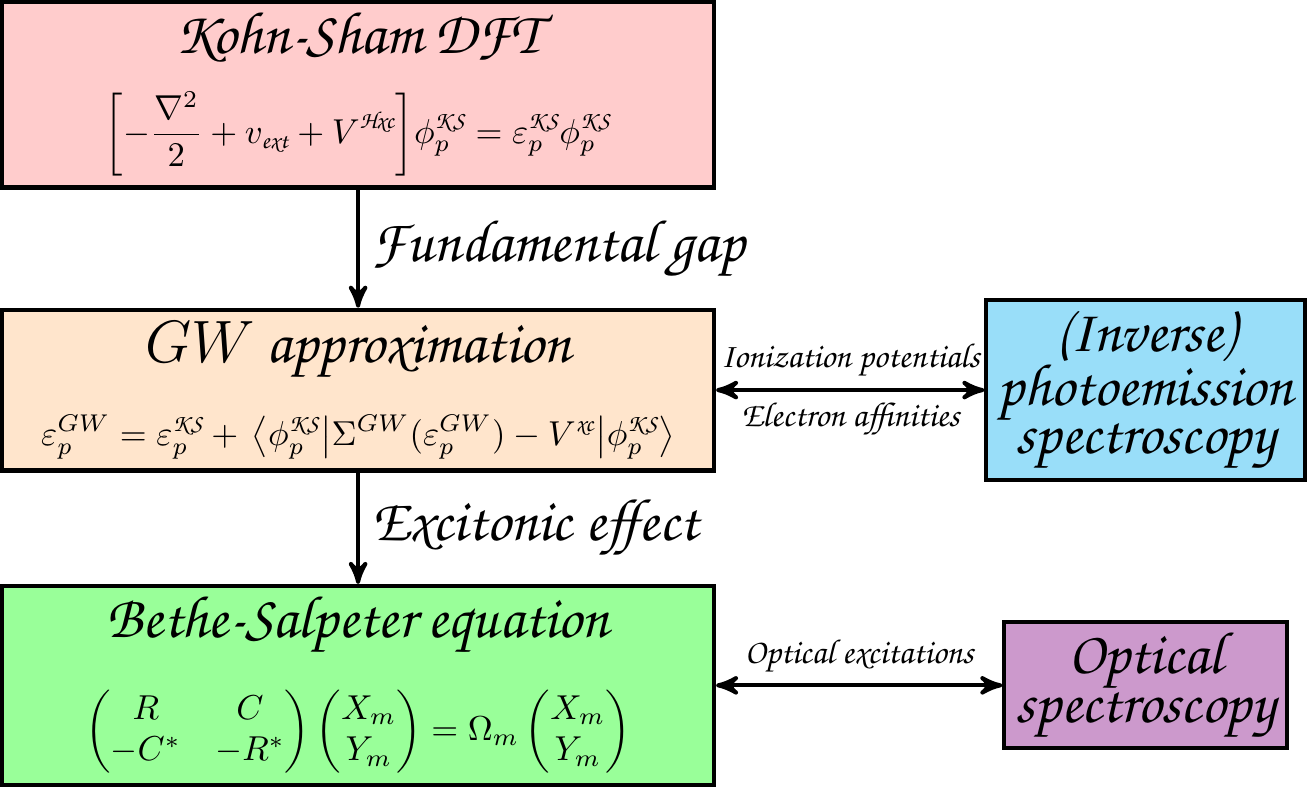}
\end{tocentry}

\begin{document}	

\ifpreprint
\else
\twocolumn[
\begin{@twocolumnfalse}
\fi
\oldmaketitle


\begin{abstract}
\titou{The Bethe-Salpeter equation (BSE) formalism is steadily asserting itself as a new efficient and accurate tool in the ensemble of computational methods available to chemists in order to predict optical excitations in molecular systems.
In particular, the combination of the so-called $GW$ approximation, giving access to reliable ionization energies and electron affinities, and the BSE formalism, able to model UV/Vis spectra, has shown to provide accurate singlet excitation energies with a typical error of $0.1$--$0.3$ eV. 
With a similar computational cost as time-dependent density-functional theory (TD-DFT), BSE is able to provide an accuracy on par with the most accurate global and range-separated hybrid functionals without the unsettling choice of the exchange-correlation functional, resolving further known issues (\textit{e.g.}, charge-transfer excitations).
In this \textit{Perspective} article, we provide a historical overview of BSE, with a particular focus on its condensed-matter roots.
We also propose a critical review of its strengths and weaknesses in different chemical situations.}
\end{abstract}

\ifpreprint
\else
\end{@twocolumnfalse}
]
\fi

\ifpreprint
\else
\small
\fi

\noindent

\paragraph{Introduction.}
In its press release announcing the attribution of the 2013 Nobel prize in Chemistry to  Karplus, Levitt, and Warshel, the Royal Swedish Academy of Sciences concluded by stating \textit{``Today the computer is just as important a tool for chemists as the test tube. 
Simulations are so realistic that they predict the outcome of traditional experiments.''} \cite{Nobel_2003} 
Martin Karplus' Nobel lecture moderated this statement, introducing his presentation by a 1929 quote from Dirac emphasizing that laws of quantum mechanics are \textit{``much too complicated to be soluble''}, urging scientists to develop \textit{``approximate practical methods''}. This is where the electronic structure community stands, attempting to develop robust approximations to study with increasing accuracy the properties of ever more complex systems. 
The study of optical excitations (also known as neutral excitations in condensed-matter systems), from molecules to extended solids, has witnessed the development of a large number of such approximate methods with numerous applications to a large variety of fields, from the prediction of the colour of precious metals for jewellery, \cite{Prandini_2019} to the understanding, \eg, of the basic principles behind organic photovoltaics, photocatalysis and DNA damage under irradiation. \cite{Kippelen_2009,Improta_2016,Wu_2019}
The present \textit{Perspective} aims at describing the current status and upcoming challenges for the Bethe-Salpeter equation (BSE) formalism \cite{Salpeter_1951,Strinati_1988} that, while sharing many features with time-dependent density-functional theory (TD-DFT), \cite{Runge_1984} including computational scaling with system size, relies on a very different formalism, with specific difficulties but also potential solutions to known TD-DFT issues. \cite{Blase_2018}
\\

\paragraph{Theory.}
The BSE formalism \cite{Salpeter_1951,Strinati_1988,Albrecht_1998,Rohlfing_1998,Benedict_1998,vanderHorst_1999} belongs to the family of  Green's function many-body perturbation theories (MBPT) \cite{Hedin_1965,Onida_2002,ReiningBook} together with, for example, the algebraic-diagrammatic construction (ADC) techniques \cite{Dreuw_2015} or the polarization propagator approaches (like SOPPA\cite{Packer_1996})  in quantum chemistry.
While the one-body density stands as the basic variable in density-functional theory (DFT), \cite{Hohenberg_1964,Kohn_1965} the pillar of Green's function MBPT is the (time-ordered) one-body Green's function
\begin{equation}
    G(\bx t,\bx't') = -i \mel{\Nel}{T \qty[ \Hat{\psi}(\bx t) \Hat{\psi}^{\dagger}(\bx't') ]}{\Nel},
\end{equation}
where $\ket{\Nel}$ is the $\Nel$-electron ground-state wave function. 
The operators $\Hat{\psi}(\bx t)$ and $\Hat{\psi}^{\dagger}(\bx't')$ remove and add an electron (respectively) in space-spin-time positions ($\bx t$) and ($\bx't'$), while $T$ is the time-ordering operator. 
For $t > t'$, $G$ provides the amplitude of probability of finding, on top of the ground-state Fermi sea (\ie, higher in energy than the highest-occupied energy level, also known as Fermi level), an electron in ($\bx t$) that was previously introduced in ($\bx't'$), while for $t < t'$ the propagation of an electron hole (often simply called a hole) is monitored. 

This definition indicates that the one-body Green's function is well suited to obtain ``charged excitations", more commonly labeled as electronic energy levels, as obtained, \eg, in a direct or inverse photo-emission experiments where an electron is ejected or added to the $N$-electron system. 
In particular, and as opposed to Kohn-Sham (KS) DFT, the Green's function formalism offers a more rigorous and systematically improvable path for the obtention of the ionization potential $I^\Nel = E_0^{\Nel-1} - E_0^\Nel$, the electronic affinity $A^\Nel = E_0^{\Nel} - E_0^{\Nel+1}$, and the experimental (photoemission) fundamental gap
\begin{equation}\label{eq:IPAEgap}
	\EgFun = I^\Nel - A^\Nel
\end{equation}
of the $\Nel$-electron system, where $E_0^\Nel$ corresponds to its ground-state energy.
Since these energy levels are key input quantities for the subsequent BSE calculation, we start by discussing these in some details.
\\

\paragraph{Charged excitations.}
A central property of the one-body Green's function is that its frequency-dependent (\ie, dynamical) spectral representation has poles at the charged excitation energies (\ie, the ionization potentials and electron affinities) of the system
\begin{equation}\label{eq:spectralG}
    G(\bx,\bx'; \omega ) = \sum_s \frac{ f_s(\bx) f^*_s(\bx') }{ \omega - \varepsilon_s + i \eta \times \text{sgn}(\varepsilon_s - \mu ) },  
\end{equation}
where $\mu$ is the chemical potential, $\eta$ is a positive infinitesimal, $\varepsilon_s = E_s^{\Nel+1} - E_0^{\Nel}$ for $\varepsilon_s > \mu$, and $\varepsilon_s = E_0^{\Nel} - E_s^{\Nel-1}$ for $\varepsilon_s < \mu$.
Here, $E_s^{\Nel}$ is the total energy of the $s$\textsuperscript{th} excited state of the $\Nel$-electron system.
The $f_s$'s  are the so-called Lehmann amplitudes that reduce to one-body orbitals in the case of single-determinant many-body wave functions (see below).
Unlike KS eigenvalues, the poles of the Green's function $\lbrace \varepsilon_s \rbrace$ are proper addition/removal energies of the $\Nel$-electron system, leading to well-defined ionization potentials and electronic affinities. 
In contrast to standard $\Delta$SCF techniques, the knowledge of $G$ provides the full ionization spectrum, as measured by direct and inverse photoemission spectroscopy, not only that associated with frontier orbitals. 

Using the equation-of-motion formalism for the creation/destruction operators, it can be shown formally that $G$ verifies
\begin{equation}\label{eq:Gmotion}
    \qty[  \pdv{}{t_1} - h(\br_1) ] G(1,2) - \int d3 \, \Sigma(1,3) G(3,2)
    = \delta(1,2),
\end{equation}
where we introduce the composite index, \eg, $1 \equiv (\bx_1 t_1)$. 
Here, $\delta$ is Dirac's delta function, $h$ is the one-body Hartree Hamiltonian and $\Sigma$ is the so-called exchange-correlation (xc) self-energy operator.
Using the spectral representation of $G$ [see Eq.~\eqref{eq:spectralG}],
dropping spin variables for simplicity, one gets the familiar eigenvalue equation, \ie,
\begin{equation}
	\titou{h(\br) f_s(\br) + \int d\br' \, \Sigma(\br,\br'; \varepsilon_s ) f_s(\br') = \varepsilon_s f_s(\br),}
\end{equation}
which formally resembles the KS equation \cite{Kohn_1965} with the difference that the self-energy $\Sigma$ is non-local, energy-dependent and non-hermitian. 
The knowledge of $\Sigma$ allows to access the true addition/removal energies, namely the entire spectrum of occupied and virtual electronic energy levels, at the cost of solving a generalized one-body eigenvalue equation. 
\\

\begin{figure}[ht]
	\includegraphics[width=0.55\linewidth]{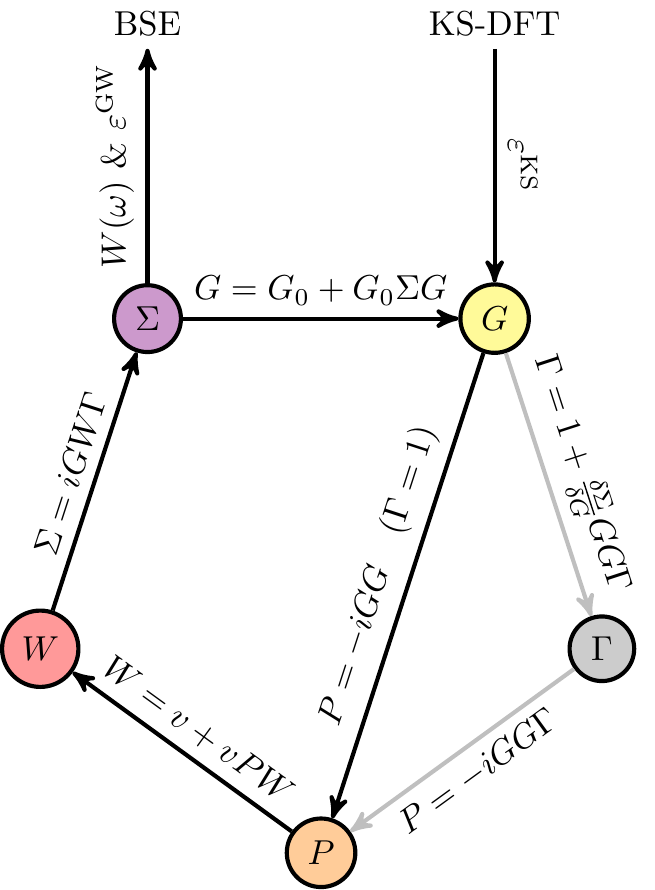}
	\caption{
	Hedin's pentagon connects the Green's function $G$, its non-interacting analog $G_0$, the irreducible vertex function $\Gamma$, the irreducible polarizability $P$, the dynamically-screened Coulomb potential $W$, and the self-energy $\Sigma$ through a set of five integro-differential equations known as Hedin's equations. \cite{Hedin_1965}
	The path made of black arrows shows the $GW$ process which bypasses the computation of $\Gamma$ (gray arrows).
	As input, one must provide KS (or HF) orbitals and their corresponding energies.
	Depending on the level of self-consistency in the $GW$ calculation, only the orbital energies or both the orbitals and their energies are corrected.
	As output, $GW$ provides corrected quantities, \ie, quasiparticle energies and $W$, which can then be used to compute the BSE optical excitations of the system of interest.
	\label{fig:pentagon}}
\end{figure}

\paragraph{The $GW$ self-energy.}
While the equations reported above are formally exact, it remains to provide an expression for the xc self-energy operator $\Sigma$.  
This is where Green's function practical theories differ. 
Developed by Lars Hedin in 1965 with application to the interacting homogeneous electron gas, \cite{Hedin_1965} the $GW$ approximation
\cite{Onida_2002,Golze_2019} follows the path of linear response by considering the variation of $G$ with respect to an external perturbation (see Fig.~\ref{fig:pentagon}). 
The resulting equation, when compared with the equation for the time-evolution of $G$ [see Eq.~\eqref{eq:Gmotion}], leads to a formal expression for the self-energy
\begin{equation}\label{eq:Sig}
	\Sigma(1,2) = i \int d34 \, G(1,4) W(3,1^{+}) \Gamma(42,3),
\end{equation}
where $W$ is the dynamically-screened Coulomb potential and $\Gamma$ is the so-called ``vertex" function. \titou{The notation $1^+$ means that the time $t_1$ is taken at $t_1^{+} = t_1 + 0^+$ for sake of causality, where $0^+$ is a positive infinitesimal.}
The neglect of the vertex, \ie, $\Gamma(42,3) = \delta(23) \delta(24)$, leads to the so-called $GW$ approximation of the self-energy 
\begin{equation}\label{eq:SigGW}
	\Sigma^{\GW}(1,2) = i \, G(1,2) W(2,1^{+}),
\end{equation}
that can be regarded as the lowest-order perturbation in terms of the screened Coulomb potential $W$ with
\begin{subequations}
\begin{gather}
	W(1,2) = v(1,2) + \int d34 \, v(1,\titou{3}) \chi_0(3,4) W(4,2),
	\label{eq:defW}
	\\
  \titou{ 	\chi_0(1,2) = -i   G(1,2^{+}) G(2,1^{+}), }
\end{gather}
\end{subequations}
where $\chi_0$ is the independent electron susceptibility and $v$ the bare Coulomb potential. 
\titou{Equation~\eqref{eq:defW} can be recast as
\begin{gather}
	W(1,2) = v(1,2) + \int d34 \, v(1,\titou{3}) \chi(3,4) v(4,2),
	\label{eq:defW2}
\end{gather}
where $\chi$ is the interacting susceptibility. In this latter expression, $(v{\chi}v)$ represents the field created in $(2)$ by the charge rearrangement of the $N$-electron system generated by a (unit) charge added in $(1)$. As such, this term contains the effect of dielectric screening (or polarization in a quantum chemist language). As in a standard $\Delta$SCF calculation, the $GW$ formalism contains the response of the $N$-electron system to an electron added (removed) to any virtual (occupied) molecular orbital, but without the restriction that only frontier orbitals can be tackled. This explains that the $GW$ one-electron energies are proper addition/removal energies.}

In practice, the input $G$ and $\chi_0$ required to initially build $\Sigma^{\GW}$ are chosen as the ``best'' Green's function and susceptibility that can be easily computed, namely the KS or Hartree-Fock (HF) ones where the $\lbrace \varepsilon_p, f_p \rbrace$ of Eq.~\eqref{eq:spectralG} are taken to be KS (or HF) eigenstates.
Taking then $( \Sigma^{\GW}-V^{\XC} )$ as a correction to the KS xc potential $V^{\XC}$, a first-order correction to the input KS energies $\lbrace \varepsilon_p^{\KS} \rbrace$ is obtained by solving the so-called quasiparticle equation
\begin{equation} \label{eq:QP-eq}
	\omega = \varepsilon_p^{\KS} +
	\mel{\phi_p^{\KS}}{\Sigma^{\GW}(\omega) - V^{\XC}}{\phi_p^{\KS}}.
\end{equation}
As a non-linear equation, the self-consistent quasiparticle equation \eqref{eq:QP-eq} has various solutions associated with different spectral weights. 
The existence of a well defined quasiparticle energy requires a solution with a large spectral weight, \ie, close to unity, a condition not always fulfilled for states far away from the fundamental gap. \cite{Veril_2018}




Such an approach, where input KS energies are corrected to yield better electronic energy levels, is labeled as the single-shot, or perturbative, $G_0W_0$ technique. 
This simple scheme was used in the early $GW$ studies of extended semiconductors and insulators, \cite{Strinati_1980,Hybertsen_1986,Godby_1988,Linden_1988} and
surfaces, \cite{Northrup_1991,Blase_1994,Rohlfing_1995} allowing to dramatically \titou{reduce} the errors associated with KS eigenvalues in conjunction with common local or gradient-corrected approximations to the xc potential.
In particular, the well-known ``band gap" problem, \cite{Perdew_1983,Sham_1983} namely the underestimation of the occupied to unoccupied bands energy gap at the local-density approximation (LDA) KS level, was dramatically reduced, bringing the agreement with experiment to within a few tenths of an eV with a computational cost scaling quartically with the system size (see below). A compilation of data for $G_0W_0$ applied to extended inorganic semiconductors can be found in Ref.~\citenum{Shishkin_2007}.

Although $G_0W_0$ provides accurate results (at least for weakly/moderately correlated systems), it is strongly starting-point dependent due to its perturbative nature. 
For example, the quasiparticle energies, and in particular the HOMO-LUMO gap, depends on the input KS eigenvalues. 
Tuning the starting point functional or applying a self-consistent $GW$ scheme are two different approaches commonly employed to tackle this problem.  
We will comment further on this particular point below when addressing the quality of the BSE optical excitations.

Another important feature compared to other perturbative techniques, the $GW$ formalism can tackle finite and periodic systems, and does not present any divergence in the limit of zero gap (metallic) systems. \cite{Campillo_1999} 
However, remaining a low-order perturbative approach starting with a single-determinant mean-field solution, it is not intended to explore strongly correlated systems. \cite{Verdozzi_1995}
\\

\begin{figure*}[ht]
	\includegraphics[width=0.7\linewidth]{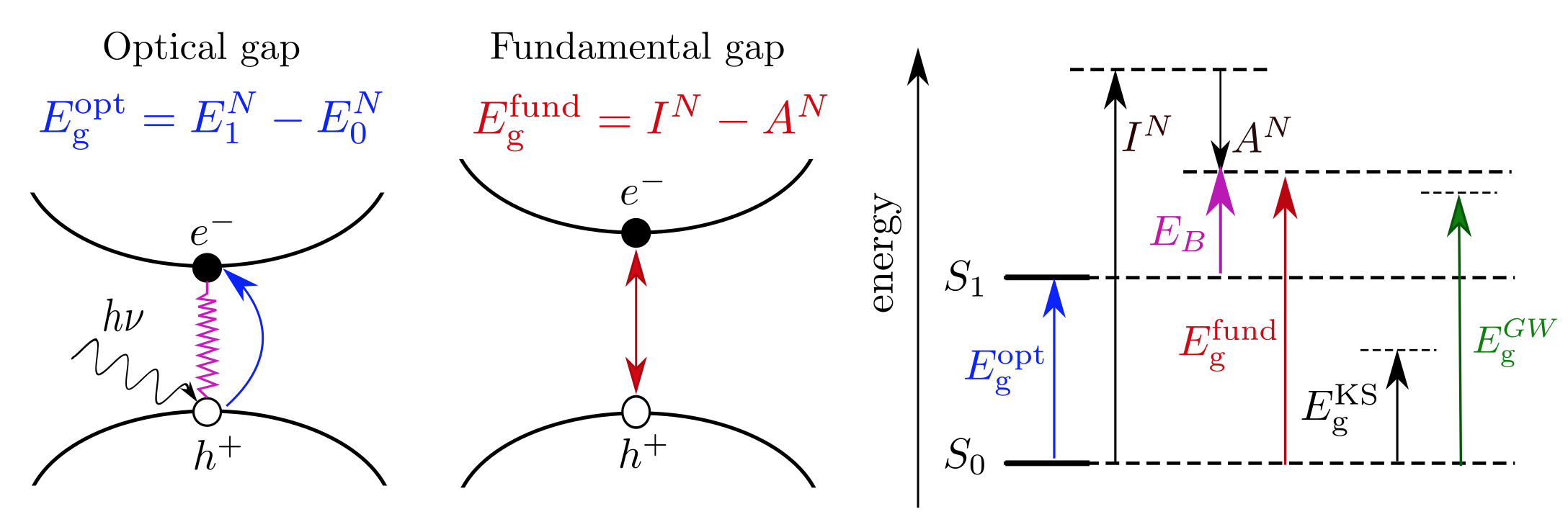}
	\caption{
	Definition of the optical gap $\EgOpt$ and fundamental gap $\EgFun$.
	$\EB$ is the electron-hole or excitonic binding energy, while $I^\Nel$ and $A^\Nel$ are the ionization potential and the electron affinity of the $\Nel$-electron system.
	$\Eg^{\KS}$ and $\Eg^{\GW}$ are the KS and $GW$ HOMO-LUMO gaps.
	See main text for the definition of the other quantities	
	\label{fig:gaps}}
\end{figure*} 

\paragraph{Neutral excitations.}
Like TD-DFT, BSE deals with the calculations of optical (or neutral) excitations, as measured by optical (\eg, absorption) spectroscopy, 
However, while TD-DFT starts with the variation of the charge density $\rho(1)$ with respect to an external local perturbation $U(1)$, the BSE formalism considers a generalized \titou{four-point} susceptibility, or two-particle correlation function, that monitors the variation of the one-body Green's function $G(1,1')$ with respect to a non-local external perturbation $U(2,2')$: \cite{Strinati_1988}
\begin{equation}
	\chi(1,2) \stackrel{\DFT}{=} \pdv{\rho(1)}{U(2)}
	\quad  \rightarrow \quad
	L(1, 2;1',2' ) \stackrel{\BSE}{=} \pdv{G(1,1')}{U(2',2)}.
\end{equation}
The formal relation $\chi(1,2) = -i L(1,2;1^+,2^+)$ with $\rho(1) = -iG(1,1^{+})$ offers a direct bridge between the TD-DFT and BSE worlds. 
The equation of motion for $G$ [see Eq.~\eqref{eq:Gmotion}] can be reformulated in the form of a Dyson equation 
\begin{equation}
	G = G_0 + G_0 ( v_H + U + \Sigma ) G,
\end{equation}
that relates the full (interacting) Green's function, $G$, to its non-interacting version, $G_0$, where $v_H$ and $U$ are the Hartree and external potentials, respectively.  
The derivative with respect to $U$ of this Dyson equation yields the self-consistent Bethe-Salpeter equation
\begin{multline}\label{eq:DysonL}
    L(1,2;1',2') = L_0(1,2;1',2') +
    \\
     \int d3456 \, L_0(1,4;1',3) \Xi^{\BSE}(3,5;4,6) L(6,2;5,2'),
\end{multline}
where $L_0(1,2;1',2') = G(1,2')G(2,1')$ is the non-interacting 4-point susceptibility and
\begin{equation}
	 i\,\Xi^{\BSE}(3,5;4,6) = v(3,6) \delta(34) \delta(56) + i \pdv{\Sigma(3,4)}{G(6,5)}
\end{equation} 
is the so-called BSE kernel.
This equation can be compared to its TD-DFT analog
\begin{equation}
    \chi(1,2) = \chi_0(1,2) + \int d34 \, \chi_0(1,3) \Xi^{\DFT}(3,4) \chi(4,2),
\end{equation}
where
\begin{equation}
    \Xi^{\DFT}(3,4) = v(3,4) + \pdv{V^{\XC}(3)}{\rho(4)}
\end{equation} 
is the TD-DFT kernel.
Plugging now the $GW$ self-energy [see Eq.~\eqref{eq:SigGW}], in a scheme that we label BSE@$GW$, leads to an approximate version of the BSE kernel
\begin{multline}\label{eq:BSEkernel}
  i\,\Xi^{\BSE}(3,5;4,6) 
  \\
  = v(3,6) \delta(34) \delta(56)  -W(3^+,4) \delta(36) \delta(45 ),
\end{multline}
where it is customary to neglect the derivative $( \partial W / \partial G)$ that introduces again higher orders in $W$. \cite{Hanke_1980,Strinati_1982,Strinati_1984}
At that stage, the BSE kernel is fully dynamical, \ie, it explicitly depends on the frequency $\omega$.
Taking the static limit, \ie, $W(\omega=0)$, for the screened Coulomb potential, that replaces the static DFT xc kernel, and expressing Eq.~\eqref{eq:DysonL} in the standard product space $\lbrace \phi_i(\br) \phi_a(\br') \rbrace$ [where $(i,j)$ are occupied spatial orbitals and $(a,b)$ are unoccupied spatial orbitals], leads to an eigenvalue problem similar to the so-called Casida equations in TD-DFT: \cite{Casida_1995}
\begin{equation} \label{eq:BSE-eigen}
    \begin{pmatrix}
		R & C  
		\\
		-C^*  & -R^{*}
	\end{pmatrix}
    \begin{pmatrix}
		X^m
		\\
		Y^m
	\end{pmatrix}
	=
	\Omega_m
    \begin{pmatrix}
		X^m  
		\\
		Y^m
	\end{pmatrix},
\end{equation}
with electron-hole ($eh$) eigenstates written as
\begin{equation}
    \psi_{m}^{eh}(\br_e,\br_h) 
    = \sum_{ia} \qty[ X_{ia}^{m} \phi_i(\br_h) \phi_a(\br_e)
	+ Y_{ia}^{m} \phi_i(\br_e) \phi_a(\br_h) ],
\end{equation}
where $m$ indexes the electronic excitations. 
The $\lbrace \phi_{i/a} \rbrace$ are typically the input (KS) eigenstates used to build the $GW$ self-energy. 
They are here taken to be real in the case of finite-size systems.
In the case of a closed-shell singlet ground state, the resonant and coupling parts of the BSE Hamiltonian read
\begin{gather}
    R_{ai,bj} = \qty( \varepsilon_a^{\GW} - \varepsilon_i^{\GW} ) \delta_{ij} \delta_{ab} + \kappa (ia|jb) - W_{ij,ab},
    \\
    C_{ai,bj} = \kappa (ia|bj) - W_{ib,aj},
\end{gather}
with $\kappa=2$ or $0$ if one targets singlet or triplet excited states (respectively), and
\begin{equation}\label{eq:Wmatel}
	W_{ij,ab} = \iint d\br d\br' 
	\phi_i(\br) \phi_j(\br) W(\br,\br'; \omega=0)
	\phi_a(\br') \phi_b(\br'),
\end{equation}
where we notice that the two occupied (virtual) eigenstates are taken at the same position of space, in contrast with the $(ia|jb)$ bare Coulomb term defined as
\begin{equation}
	(ia|jb) = \iint d\br d\br' 
	\phi_i(\br) \phi_a(\br) v(\br-\br')
	\phi_j(\br') \phi_b(\br').
\end{equation}  
Neglecting the coupling term $C$ between the resonant term $R$ and anti-resonant term $-R^*$ in Eq.~\eqref{eq:BSE-eigen}, leads to the well-known Tamm-Dancoff approximation (TDA).

As compared to TD-DFT: i) the $GW$ quasiparticle energies $\lbrace \varepsilon_{i/a}^{\GW} \rbrace$ replace the KS eigenvalues, and ii) the non-local screened Coulomb matrix elements replaces the DFT xc kernel.
We emphasize that these equations can be solved at exactly the same cost as the standard TD-DFT equations once the quasiparticle energies and screened Coulomb potential $W$ are inherited from preceding $GW$ calculations. 
This defines the standard (static) BSE@$GW$ scheme that we discuss in this \textit{Perspective}, highlighting its \emph{pros} and \emph{cons}.  
\\

\paragraph{Historical overview.}
Originally developed in the framework of nuclear physics, \cite{Salpeter_1951} the BSE formalism has emerged in condensed-matter physics around the 1960's at the tight-binding level with the study of the optical properties of simple semiconductors. \cite{Sham_1966,Strinati_1984,Delerue_2000}
Three decades later, the first \textit{ab initio} implementations, starting with small clusters, \cite{Onida_1995,Rohlfing_1998} extended semiconductors, and wide-gap insulators \cite{Albrecht_1997,Benedict_1998,Rohlfing_1999b} paved the way to the popularization in the solid-state physics community of the BSE formalism. 

Following pioneering applications to periodic polymers and molecules, \cite{Rohlfing_1999a,Horst_1999,Puschnig_2002,Tiago_2003} BSE gained much momentum in quantum chemistry \cite{listofrefs} with, in particular, several benchmarks \cite{Boulanger_2014,Jacquemin_2015a,Bruneval_2015,Jacquemin_2015b,Hirose_2015,Jacquemin_2017,Krause_2017,Gui_2018} on large molecular sets performed with the very same parameters (geometries, basis sets, etc) than the available higher-level reference calculations. \cite{Schreiber_2008} 
Such comparisons were grounded in the development of codes replacing the plane-wave paradigm of solid-state physics by Gaussian basis sets, together with adequate auxiliary bases when resolution-of-the-identity (RI) techniques \cite{Ren_2012b} were used.  

An important conclusion drawn from these calculations was that the quality of the BSE excitation energies is strongly correlated to the deviation of the preceding $GW$ HOMO-LUMO gap 
\begin{equation}
	\Eg^{\GW} = \eps_{\LUMO}^{\GW} - \varepsilon_{\HOMO}^{\GW},
\end{equation}
with the experimental (photoemission) fundamental gap defined in Eq.~\eqref{eq:IPAEgap}.

Standard $G_0W_0$ calculations starting with KS eigenstates generated with (semi)local functionals yield much larger HOMO-LUMO gaps than the input KS gap
\begin{equation}
	\Eg^{\KS} = \eps_{\LUMO}^{\KS} - \varepsilon_{\HOMO}^{\KS},
\end{equation}
but still too small as compared to the experimental value, \ie,
\begin{equation}
	\Eg^{\KS} \ll \Eg^{G_0W_0} < \EgFun.
\end{equation}
\titou{Such a residual discrepancy has been attributed by several authors to ``overscreening", namely the effect associated with building the susceptibility $\chi$ based on a grossly underestimated (KS) band gap. This leads to a spurious enhancement of the screening or polarization and, consequently, to an underestimated $G_0W_0$ gap as compared to the (exact) fundamental gap. More prosaically, the $G_0W_0$ approach is constructed as a first-order perturbation theory, so by correcting a very ``bad" zeroth-order KS system one cannot expect to obtain an accurate corrected gap.}
Such an underestimation of the fundamental gap leads to a similar underestimation of the optical gap $\EgOpt$, \ie, the lowest optical excitation energy:
\begin{equation}
	\EgOpt = E_1^{\Nel} - E_0^{\Nel} = \EgFun + \EB,
\end{equation}
where $\EB$ accounts for the excitonic effect, that is, the stabilization induced by the attraction of the excited electron and its hole left behind (see Fig.~\ref{fig:gaps}).

Such a residual gap problem can be significantly improved by adopting xc functionals with a tuned amount of exact exchange \cite{Stein_2009,Kronik_2012} that yield a much improved KS gap as a starting point for the $GW$ correction. \cite{Bruneval_2013,Rangel_2016,Knight_2016,Gui_2018} 
Alternatively, self-consistent approaches such as eigenvalue self-consistent (ev$GW$) \cite{Hybertsen_1986} or quasiparticle self-consistent (qs$GW$) \cite{vanSchilfgaarde_2006} schemes, where corrected eigenvalues, and possibly orbitals, are reinjected in the construction of $G$ and $W$, have been shown to lead to a significant improvement of the quasiparticle energies in the case of molecular systems, with the advantage of significantly removing the dependence on the starting point functional. \cite{Rostgaard_2010,Blase_2011,Ke_2011,Rangel_2016,Kaplan_2016,Caruso_2016} 
As a result, BSE singlet excitation energies starting from such improved quasiparticle energies were found to be in much better agreement with reference calculations.
For sake of illustration, an average error of $0.2$ eV was found for the well-known Thiel set \cite{Schreiber_2008} gathering ca.~200 representative singlet excitations from a large variety of representative molecules. \cite{Jacquemin_2015a,Bruneval_2015,Gui_2018,Krause_2017} 
This is equivalent to the best TD-DFT results obtained by scanning a large variety of hybrid functionals with various amounts of exact exchange. 

\begin{figure}[ht]
	\includegraphics[width=0.6\linewidth]{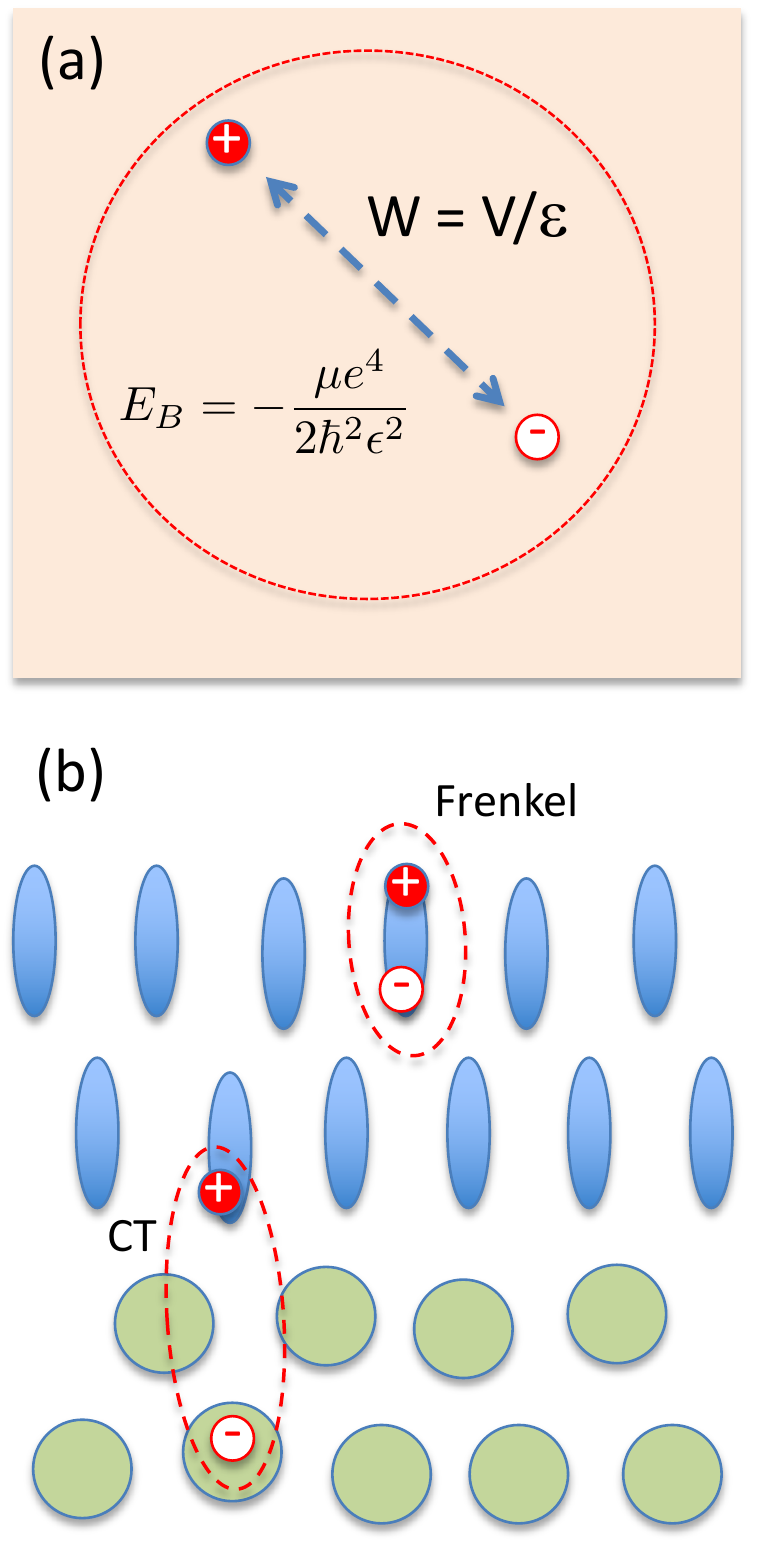}
	\caption{
		Symbolic representation of (a) extended Wannier exciton with large electron-hole average distance, and (b) Frenkel (local) and charge-transfer (CT) excitations at a donor-acceptor interface. 
		Wannier and CT excitations require long-range electron-hole interaction accounting for the host dielectric constant. 
		In the case of Wannier excitons, the binding energy $\EB$ can be well approximated by the standard hydrogenoid model where $\mu$ is the effective mass and $\epsilon$ is the dielectric constant.
	\label{fig:CTfig}}
\end{figure}

\paragraph{Charge-transfer excited states.} 
A very remarkable success of the BSE formalism lies in the description of charge-transfer (CT) excitations, a notoriously difficult problem for TD-DFT adopting standard (semi-)local functionals. \cite{Dreuw_2004} 
Similar difficulties emerge in solid-state physics for semiconductors where extended Wannier excitons, characterized by weakly overlapping electrons and holes (Fig.~\ref{fig:CTfig}), cause a dramatic deficit of spectral weight at low energy. \cite{Botti_2004} 
These difficulties can be ascribed to the lack of long-range electron-hole interaction with local xc functionals.
It can be cured through an exact exchange contribution, a solution that explains the success of (optimally-tuned) range-separated hybrids for the description of CT excitations. \cite{Stein_2009,Kronik_2012} 
The analysis of the screened Coulomb potential matrix elements in the BSE kernel [see Eq.~\eqref{eq:BSEkernel}] reveals that such long-range (non-local) electron-hole interactions are properly described, including in environments (solvents, molecular solid, etc.) where the screening reduces the long-range electron-hole interactions. 
The success of the BSE formalism to treat CT excitations has been demonstrated in several studies, \cite{Rocca_2010,Cudazzo_2010,Lastra_2011,Blase_2011,Baumeier_2012a,Duchemin_2012,Sharifzadeh_2013,Cudazzo_2013,Yin_2014} opening the way to the modeling of key applications such as doping, \cite{Li_2017b} photovoltaics or photocatalysis in organic systems.\\

\paragraph{Combining BSE with PCM and QM/MM models.} 
The ability to account for the effect on the excitation energies of an electrostatic and dielectric environment (an electrode, a solvent, a molecular interface\ldots) is an important step towards the description of realistic systems. 
Pioneering BSE studies demonstrated, for example, the large renormalization of charged and neutral excitations in molecular systems and nanotubes close to a metallic electrode or in bundles. \cite{Lastra_2011,Rohlfing_2012,Spataru_2013}
Recent attempts to merge the $GW$ and BSE formalisms with model polarizable environments at the PCM or QM/MM levels
\cite{Baumeier_2014,Duchemin_2016,Li_2016,Varsano_2016,Duchemin_2018,Li_2019,Tirimbo_2020} paved the way not only to interesting applications but also to a better understanding of the merits of these approaches relying on the use of the screened Coulomb potential designed to capture polarization effects at all spatial ranges. 
As a matter of fact, dressing the bare Coulomb potential with the reaction field matrix 
$[
v(\br,\br') \longrightarrow v(\br,\br') + v^{\text{reac}}(\br,\br'; \omega)
]$
in the relation between the screened Coulomb potential $W$ and the independent-electron susceptibility [see Eq.~\eqref{eq:defW}] allows to perform $GW$ and BSE calculations in a polarizable environment at the same computational cost as the corresponding gas-phase calculation. 
The reaction field operator $v^{\text{reac}}(\br,\br'; \omega)$ describes the potential generated in $\br'$ by the charge rearrangements in the polarizable environment induced by a source charge located in $\br$, where $\br$ and $\br'$ lie in the quantum mechanical subsystem of interest. 
The reaction field is dynamical since the dielectric properties of the environment, such as the macroscopic dielectric constant $\epsilon_M(\omega)$, are in principle frequency dependent.   
Once the reaction field matrix is known, with typically $\order*{\Norb N_\text{MM}^2}$ operations (where $\Norb$ is the number of orbitals and $N_\text{MM}$ the number of polarizable atoms in the environment), the full spectrum of $GW$ quasiparticle energies and BSE neutral excitations can be renormalized by the effect of the environment.

A remarkable property \cite{Duchemin_2018} of the scheme described above, which combines the BSE formalism with a polarizable environment, is that the renormalization of the electron-electron and electron-hole interactions by the reaction field captures both linear-response and state-specific contributions \cite{Cammi_2005} to the solvatochromic shift of the optical lines, allowing to treat on the same footing local (Frenkel) and CT excitations. 
This is an important advantage as compared to, \eg, TD-DFT where linear-response and state-specific effects have to be explored with different formalisms. 

To date, environmental effects on fast electronic excitations are only included by considering the low-frequency optical response of the polarizable medium (\eg, considering the $\epsilon_{\infty} \simeq 1.78$ macroscopic dielectric constant of water in the optical range), neglecting the frequency dependence of the dielectric constant in the optical range. 
Generalization to fully frequency-dependent polarizable properties of the environment would allow to explore systems where the relative dynamics of the solute and the solvent are not decoupled, \ie, situations where neither the adiabatic limit nor the anti-adiabatic limits are expected to be valid (for a recent discussion, see Ref.
~\citenum{Huu_2020}).

We now leave the description of successes to discuss difficulties and future directions of developments and improvements.
\\

\paragraph{The computational challenge.} 
As emphasized above, the BSE eigenvalue equation in the single-excitation space [see Eq.~\eqref{eq:BSE-eigen}] is formally equivalent to that of TD-DFT or TD-HF. \cite{Dreuw_2005}
Searching iteratively for the lowest eigenstates exhibits the same $\order*{\Norb^4}$ matrix-vector multiplication computational cost within BSE and TD-DFT. 
Concerning the construction of the BSE Hamiltonian, it is no more expensive than building its TD-DFT analogue with hybrid functionals, reducing again to $\order*{\Norb^4}$ operations with standard RI techniques. 
Explicit calculation of the full BSE Hamiltonian in transition space can be further avoided using density matrix perturbation theory, \cite{Rocca_2010,Nguyen_2019} not reducing though the $\order*{\Norb^4}$ scaling, but sacrificing further the knowledge of the eigenvectors. 
Exploiting further the locality of the atomic orbital basis, the BSE absorption spectrum can be obtained with $\order*{\Norb^3}$ operations using such iterative techniques. \cite{Ljungberg_2015} 
With the same restriction on the eigenvectors, a time-propagation approach, similar to that implemented for TD-DFT, \cite{Yabana_1996} combined with stochastic techniques to reduce the cost of building the BSE Hamiltonian matrix elements, allows quadratic scaling with systems size. \cite{Rabani_2015}

In practice, the main bottleneck for standard BSE calculations as compared to TD-DFT resides in the preceding $GW$ calculation that scales as $\order{\Norb^4}$ with system size using plane-wave basis sets or RI techniques, but with a rather large prefactor. 
The field of low-scaling $GW$ calculations is however witnessing significant advances. 
While the sparsity of, for example, the overlap matrix in the atomic orbital basis allows to reduce the scaling in the large size limit, \cite{Foerster_2011,Wilhelm_2018} efficient real-space grids and time techniques are blooming, \cite{Rojas_1995,Liu_2016} borrowing in particular the well-known Laplace transform approach used in quantum chemistry. \cite{Haser_1992}
Together with a stochastic sampling of virtual states, this family of techniques allow to set up linear scaling $GW$ calculations. \cite{Vlcek_2017} 
The separability of occupied and virtual states summations lying at the heart of these approaches are now spreading fast in quantum chemistry within the interpolative separable density fitting (ISDF) approach applied for calculating with cubic scaling the susceptibility needed in random-phase approximation (RPA) and $GW$ calculations. \cite{Lu_2017,Duchemin_2019,Gao_2020}  
These ongoing developments pave the way to applying the $GW$@BSE formalism to systems containing several hundred atoms on standard laboratory clusters.
\\

\paragraph{The triplet instability challenge.}
The analysis of the singlet-triplet splitting is central to numerous applications such as singlet fission or thermally activated delayed fluorescence (TADF).
From a more theoretical point of view, triplet instabilities that often \titou{plague} the applicability of TD-DFT are intimately linked to the stability analysis of restricted closed-shell solutions at the HF \cite{Seeger_1977} and KS \cite{Bauernschmitt_1996} levels.
While TD-DFT with range-separated hybrids can benefit from tuning the range-separation parameter(s) as a mean to act on the triplet instability, \cite{Sears_2011} BSE calculations do not offer this pragmatic way-out since the screened Coulomb potential that builds the kernel does not offer any parameter to tune.

Benchmark calculations \cite{Jacquemin_2017b,Rangel_2017} clearly concluded that triplets are notably too low in energy within BSE and that the use of the TDA was able to partly reduce this error. 
However, as it stands, the BSE accuracy for triplets remains rather unsatisfactory for reliable applications. 
An alternative cure was offered by hybridizing TD-DFT and BSE, that is, by adding to the BSE kernel the correlation part of the underlying DFT functional used to build the susceptibility and resulting screened Coulomb potential $W$. \cite{Holzer_2018a}
\\

\paragraph{The challenge of the ground-state energy.}
In contrast to TD-DFT which relies on KS-DFT as its ground-state analog, the ground-state BSE energy is not a well-defined quantity, and no clear consensus has been found regarding its formal definition.
Consequently, the BSE ground-state formalism remains in its infancy with very few available studies for atomic and molecular systems. \cite{Olsen_2014,Holzer_2018b,Li_2019,Li_2020,Loos_2020}

A promising route, which closely follows RPA-type formalisms, \cite{Angyan_2011} is to calculate the ground-state BSE energy within the adiabatic-connection fluctuation-dissipation theorem (ACFDT) framework. \cite{Furche_2005}
Thanks to comparisons with both similar and state-of-art computational approaches, it was recently shown that the ACFDT@BSE@$GW$ approach yields extremely accurate PES around equilibrium, and can even compete with high-order coupled cluster methods in terms of absolute ground-state energies and equilibrium distances. \cite{Loos_2020}
However, their accuracy near the dissociation limit remains an open question. \cite{Caruso_2013,Olsen_2014,Colonna_2014,Hellgren_2015,Holzer_2018b}
Indeed, in the largest available benchmark study \cite{Holzer_2018b} encompassing the total energies of the atoms \ce{H}--\ce{Ne}, the atomization energies of the 26 small molecules forming the HEAT test set, and the bond lengths and harmonic vibrational frequencies of $3d$ transition-metal monoxides, the BSE correlation energy, as evaluated within the ACFDT framework, \cite{Furche_2005} was mostly discarded from the set of tested techniques due to instabilities (negative frequency modes in the BSE polarization propagator) and replaced by an approximate (RPAsX) approach where the screened-Coulomb potential matrix elements was removed from the resonant electron-hole contribution. \cite{Maggio_2016,Holzer_2018b} 
Moreover, it was also observed in Ref.~\citenum{Loos_2020} that, in some cases, unphysical irregularities on the ground-state PES appear  due to the appearance of discontinuities as a function of the bond length for some of the $GW$ quasiparticle energies. 
Such an unphysical behavior stems from defining the quasiparticle energy as the solution of the quasiparticle equation with the largest spectral weight in cases where several solutions can be found [see Eq.~\eqref{eq:QP-eq}].
We refer the interested reader to Refs.~\citenum{vanSetten_2015,Maggio_2017,Loos_2018,Veril_2018,Duchemin_2020} for detailed discussions.
\\

\paragraph{The challenge of analytical nuclear gradients.}
The features of ground- and excited-state potential energy surfaces (PES) are critical for the faithful description and a deeper understanding of photochemical and photophysical processes. \cite{Olivucci_2010}
For example, chemoluminescence and fluorescence are associated with geometric relaxation of excited states, and structural changes upon electronic excitation. \cite{Navizet_2011}
Reliable predictions of these mechanisms, which have attracted much experimental and theoretical interest lately, require exploring the ground- and excited-state PES. 
From a theoretical point of view, the accurate prediction of excited electronic states remains a challenge, \cite{Loos_2020a} especially for large systems where state-of-the-art computational techniques (such as multiconfigurational methods \cite{Roos_1996}) cannot be afforded.
For the last two decades, TD-DFT has been the go-to method to compute absorption and emission spectra in large molecular systems.

In TD-DFT, the PES for the excited states can be easily and efficiently obtained as a function of the molecular geometry by simply adding the ground-state DFT energy to the excitation energy of the selected state. 
One of the strongest assets of TD-DFT is the availability of first- and second-order analytic nuclear gradients (\ie, the first- and second-order derivatives of the excited-state energy with respect to atomic displacements), which enables the exploration of excited-state PES. \cite{Furche_2002} 

A significant limitation of the BSE formalism, as compared to TD-DFT, lies in the lack of analytical nuclear gradients for both the ground and excited states, preventing efficient studies of many key excited-state processes.
While calculations of the $GW$ quasiparticle energy ionic gradients is becoming increasingly popular,
\cite{Lazzeri_2008,Faber_2011b,Yin_2013,Montserrat_2016,Zhenglu_2019} only one pioneering study of the excited-state BSE gradients has been published so far. \cite{Beigi_2003} 
In this seminal work devoted to small molecules (\ce{CO} and \ce{NH3}), only the BSE excitation energy gradients were calculated, with the approximation that the gradient of the screened Coulomb potential can be neglected, computing further the KS-LDA forces as its ground-state contribution.
\\

\paragraph{Beyond the static approximation.}
Going beyond the static approximation is a difficult challenge which has been, nonetheless, embraced by several groups.\cite{Strinati_1988,Rohlfing_2000,Ma_2009a,Ma_2009b,Romaniello_2009b,Sangalli_2011,Huix-Rotllant_2011,Zhang_2013,Rebolini_2016,Olevano_2019}
As mentioned earlier in this \textit{Perspective}, most of BSE calculations are performed within the so-called static approximation, which substitutes the dynamically-screened (\ie, frequency-dependent) Coulomb potential $W(\omega)$ by its static limit $W(\omega = 0)$ [see Eq.~\eqref{eq:Wmatel}].
It is important to mention that diagonalizing the BSE Hamiltonian in the static approximation corresponds to solving a \textit{linear} eigenvalue problem in the space of single excitations, while it is, in its dynamical form, a non-linear eigenvalue problem (in the same space) which is much harder to solve from a numerical point of view.
In complete analogy with the ubiquitous adiabatic approximation in TD-DFT, one key consequence of the static approximation is that double (and higher) excitations are completely absent from the BSE optical spectrum, which obviously hampers the applicability of BSE as \titou{double excitations} may play, indirectly, a key role in photochemistry mechanisms. 
Higher excitations would be explicitly present in the BSE Hamiltonian by ``unfolding'' the dynamical BSE kernel, and one would recover a linear eigenvalue problem with, nonetheless, a much larger dimension.
Corrections to take into account the dynamical nature of the screening may or may not recover these multiple excitations.
However, dynamical corrections permit, in any case, to recover, for transitions with a dominant single-excitation character, additional relaxation effects coming from higher excitations.

From a more practical point of view, dynamical effects have been found to affect the positions and widths of core-exciton resonances in semiconductors, \cite{Strinati_1982,Strinati_1984} rare gas solids, and transition metals. \cite{Ankudinov_2003}
Thanks to first-order perturbation theory, Rohlfing and coworkers have developed an efficient way of taking into account the dynamical effects via a plasmon-pole approximation combined with TDA. \cite{Rohlfing_2000,Ma_2009a,Ma_2009b,Baumeier_2012b}
With such a scheme, they have been able to compute the excited states of biological chromophores, showing that taking into account the electron-hole dynamical screening is important for an accurate description of the lowest $n \ra \pi^*$ excitations. \cite{Ma_2009a,Ma_2009b,Baumeier_2012b}
Studying PYP, retinal and GFP chromophore models, Ma \textit{et al.}~found that \textit{``the influence of dynamical screening on the excitation energies is about $0.1$ eV for the lowest $\pi \ra \pi^*$ transitions, but for the lowest $n \ra \pi^*$ transitions the influence is larger, up to $0.25$ eV.''} \cite{Ma_2009b}
Zhang \textit{et al.}~have studied the frequency-dependent second-order BSE kernel and they have observed an appreciable improvement over configuration interaction with singles (CIS), time-dependent Hartree-Fock (TDHF), and adiabatic TD-DFT results. \cite{Zhang_2013}
Rebolini and Toulouse have performed a similar investigation in a range-separated context, and they have reported a modest improvement over its static counterpart. \cite{Rebolini_2016} 
In these two latter studies, they also followed a (non-self-consistent) perturbative approach within TDA with a renormalization of the first-order perturbative correction.
\\

\paragraph{Conclusion.}
Although far from being exhaustive, we hope that this \textit{Perspective} provides a concise and fair assessment of the strengths and weaknesses of the BSE formalism of many-body perturbation theory.
To do so, we have briefly reviewed the theoretical aspects behind BSE, and its intimate link with the underlying $GW$ calculation that one must perform to compute quasiparticle energies and the dynamically-screened Coulomb potential; two of the key input ingredients associated with the BSE formalism.
We have then provided a succinct historical overview with a particular focus on its condensed-matter roots, and the lessons that the community has learnt from several systematic benchmark studies on large molecular systems.
Several success stories are then discussed (charge-transfer excited states and combination with reaction field methods), before debating some of the challenges faced by the BSE formalism (computational cost, triplet instabilities, ambiguity in the definition of the ground-state energy, lack of analytical nuclear gradients, and limitations due to the static approximation). 
We hope that, by providing a snapshot of the ability of BSE in 2020, the present \textit{Perspective} article will motivate a larger community to participate to the development of this alternative to TD-DFT which, we believe, may become a very valuable computational tool for the physical chemistry community.
\\

\section*{Acknowledgments}
PFL thanks the European Research Council (ERC) under the European Union's Horizon 2020 research and innovation programme (Grant agreement No.~863481) for financial support.
Funding from the \textit{``Centre National de la Recherche Scientifique''} is also acknowledged. 
This work has also been supported through the EUR Grant NanoX ANR-17-EURE-0009 in the framework of the \textit{``Programme des Investissements d'Avenir''}.
DJ acknowledges the \textit{R\'egion des Pays de la Loire} for financial support. \\

\bibliography{BSE_JPCL}

\end{document}